\documentclass[aps,floatfix,preprint,showpacs,preprintnumbers,nofootinbib,superscriptaddress,natbib]{revtex4}

\usepackage{graphicx,float}
\usepackage{epsfig}		
\usepackage{dcolumn}
\usepackage{bm}
\usepackage{amsmath,upgreek}
\usepackage{amssymb}
\usepackage{amssymb}
\usepackage{braket}
\usepackage{natbib}

\usepackage[all]{xy}
\usepackage{pdfpages}
\usepackage{color}
\usepackage[colorlinks=true,
            linkcolor=red,
            urlcolor=purple,
            citecolor=blue,hyperindex]{hyperref}

\def\0{\mbox{\tiny $0$}}
\def\1{\mbox{\tiny $1$}}
\def\2{\mbox{\tiny $2$}}
\def\3{\mbox{\tiny $3$}}
\def\4{\mbox{\tiny $4$}}
\def\5{\mbox{\tiny $5$}}
\def\6{\mbox{\tiny $6$}}
\def\7{\mbox{\tiny $7$}}
\def\8{\mbox{\tiny $8$}}
\def\9{\mbox{\tiny $9$}}

\def\f14{\mbox{\tiny $\frac{1}{4}$}}

\newcommand*{\defeq}{\mathrel{\vcenter{\baselineskip0.5ex \lineskiplimit0pt
                     \hbox{\scriptsize.}\hbox{\scriptsize.}}}%
                     =}

\begin{document}

\title{Collapsing Shells and Black Holes: a quantum analysis}

\author{P. Leal}
\email{up201002687@fc.up.pt}
\affiliation{Departamento de F\'isica e Astronomia and Centro de F\'isica do Porto, Faculdade de Ci\^{e}ncias da
Universidade do Porto, Rua do Campo Alegre 687, 4169-007, Porto, Portugal.}
\author{A. E. Bernardini}
\email{alexeb@ufscar.br}
\affiliation{Departamento de F\'isica e Astronomia and Centro de F\'isica do Porto, Faculdade de Ci\^{e}ncias da
Universidade do Porto, Rua do Campo Alegre 687, 4169-007, Porto, Portugal.}
\altaffiliation[On leave of absence from]{~Departamento de F\'{\i}sica, Universidade Federal de S\~ao Carlos, PO Box 676, 13565-905, S\~ao Carlos, SP, Brasil.}
\author{O. Bertolami}
\email{orfeu.bertolami@fc.up.pt}
\affiliation{Departamento de F\'isica e Astronomia and Centro de F\'isica do Porto, Faculdade de Ci\^{e}ncias da
Universidade do Porto, Rua do Campo Alegre 687, 4169-007, Porto, Portugal.}


\begin{abstract}
The quantization of a spherically symmetric null shells is performed and extended to the framework of phase-space noncommutative (NC) quantum mechanics.
The encountered properties are investigated making use of the Israel junction conditions on the shell, considering that it is the boundary between two spherically symmetric spacetimes.
Using this method, and considering two different Kantowski-Sachs spacetimes as a representation for the Schwarzschild spacetime, the relevant quantities on the shell are computed, such as its stress-energy tensor and the action for the whole spacetime. From the obtained action, the Wheeler-deWitt equation is deduced in order to provide the quantum framework for the system. Solutions for the wavefunction of the system are found on both the commutative and NC scenarios. It is shown that, on the commutative version, the wave function has a purely oscillatory behavior in the interior of the shell. In the NC setting, it is shown that the wavefunction vanishes at the singularity, as well as, at the event horizon of the black hole.
\end{abstract}

\date{\today}
\maketitle

\section{Introduction}




It is well-known that General Relativity (GR), as a description for gravity, is not compatible with the quantum formalism that governs the remaining fundamental interactions of Nature. Therefore, a quantum theory of gravity is needed. One can attempt to quantize gravity directly from GR, in a procedure similar to those ones to which classical field theories have undergone. This treatment is realized through the ADM formalism \cite{ADM_1} for GR, through a canonical quantization procedure that leads to the Wheeler-deWitt (WdW) equation \cite{Dewitt_1967,Wheeler_1968}. This method gives origin to a functional differential equation for the metric components which, in the general case, does not exhibit neither analytical nor stable numerical solutions. Furthermore, the interpretation of such a functional framework and its comparison with the classical theory are not trivial. However, for a special class of metrics, the functional equations become differential equations, bearing much more tractable mathematics, and a Klein-Gordon like equation arises for the metric components. This simplification, often referred to as the minisuperspace approximation \cite{Wheeler_1968,Gravitation_1973,Vilenkin_1985,Hawking_1983}, provides a simple framework for quantizing some interesting classes of metrics.
This method has been used to study the quantum dynamics of several geometries, like ones with $SO(3)$ invariance and cosmological models \cite{Vilenkin_1985,Hawking_1983,Bertolami_1991,Bertolami_1995,Bertolami_1997,Halliwell_1985}. The quantum cosmology of the Kantwosky-Sachs (KS) metric was studied in the context of noncommutative quantum mechanics in Refs. \cite{Catarina_2008,Catarina_2008_2}. For the BH problem, a map from the Schwarzschild to a KS metric is considered. This allows for the introduction of conjugate momenta for the metric components in a spherically symmetric metric so to yield a Hamiltonian, that is a sum of constraints, and consequently, a well defined WdW equation for the problem \cite{Compean_2002,Catarina_2010,Catarina_2010_2}. In this context, one of the purposes of the BH quantization followed along this manuscript (c. f. Sec.~\ref{map_section}) is to examine the collapse of a shell into a BH.

In the setting to be studied, the gravitational collapse, namely, the collapse of a thin shell, which acts as a boundary dividing two spacetimes, is the most natural problem to be addressed. The standard method for solving this class of problems was introduced by Israel \cite{Israel_1966} in his formulation of the junction conditions needed for the resulting spacetime to be compatible with Einstein field equations. Although it was first formulated for timelike and spacelike shells, it was later extended for the case of null shells \cite{Barrabes_1991,Clarke_1987}. The use of this formalism is two-fold. On one side, it is possible to define the energy-momentum tensor on the shell and deduce what are the possible spacetimes to be matched: the choice of one of them completely determines the features of the other. On the other side, one can specify the spacetimes to be connected and use the junction conditions to define relevant the quantities on the shell. Most of the literature on the quantization of collapsing shells is only concerned with the first approach (see, for instance, Refs. \cite{Louko_1998,Kiefer_2015} and references therein). In this work the second approach is chosen so to allow for a straightforward definition of the shell quantities in terms of the metric function of the embedding spacetimes.

Upon quantization, a set of commutation relations must be imposed on the metric components and their conjugate momenta. Usually these relations follow directly from the Poisson brackets of the corresponding quantities at classical level, and can be implemented at quantum level through the Heisenberg-Weyl algebra. Interestingly, these commutation relations can be deformed so that the parameter space becomes noncommutative (NC), as discussed at quantum mechanical level, e.g, in Refs. \cite{Catarina_2008,Catarina_2008_2,2013,Bastos001,Belissard_1994,Rosa_2005,Nair_2001,Duval_2001,Jian_2004_1,Jian_2004_2,Demetrian,Gamboa_2001}.
The application to the deformed quantization of GR solutions via WdW equation has been considered for the KS metric \cite{Catarina_2008} and BH \cite{Catarina_2010,Catarina_2010_2}. It was shown that the phase-sapce NC deformation of the algebra works as a regulator for the BH singularity at the origin \cite{Catarina_2010_2}. It is, thus, interesting to study the effect of the deformed algebra in the context of the gravitational collapsing shells.

This paper is structured as follows. In Sec.~\ref{map_section}, the GR geometrical construction is performed as to explicitly express the metrics on both sides of the shell. A convenient map parametrization is introduced to simplify the ensuing developments. In Sec.~\ref{action_section}, the action for the spacetime with a shell is computed considering a discontinuity term in the Ricci scalar and a boundary term for null-like surfaces. In addition, the Hamiltonian for the system is deduced. In Sec.~\ref{quantization_section}, the Wheeler-deWitt quantization procedure is adopted and the Hamiltonian constraint equation is solved in commutative and phase-space NC scenarios. Our conclusions are drawn in Sec.~\ref{conclusions_section}.

\section{The Black hole-mass shell} \label{map_section}

Let us consider a null shell of mass, $m$, falling into a Schwarzschild BH of mass, $M$. There are two metrics in this system, one corresponding to the interior of the shell and the other related to the exterior. These can be written as:
\begin{subequations} \label{initialmetrics}
\begin{equation}
\mathrm{d}s^2_{in}=-\left(1-\frac{2M}{r}\right)\mathrm{d}t^2+\left(1-\frac{2M}{r}\right)^{-1}\mathrm{d}r^2+r^2\left(\mathrm{d}\theta^2+\sin^2\theta\,\mathrm{d}\phi^2\right),
\end{equation}
\begin{equation}
\mathrm{d}s^2_{out}=-\left(1-\frac{2(M+m)}{r}\right)\mathrm{d}t^2+\left(1-\frac{2(M+m)}{r}\right)^{-1}\mathrm{d}r^2+r^2\left(\mathrm{d}\theta^2+\sin^2\theta\,\mathrm{d}\phi^2\right).
\end{equation}
\end{subequations}
Depending on the position of the shell, there can be two event horizons: one at $r=2M$ and the other at $r=2(m+M)$. Since one intends to analyze the behavior of the shell as it collapses into the interior region, in this work we will consider the shell located at $r<2M$. This choice makes the temporal and radial components of both metrics in Eq.~(\ref{initialmetrics}) negative and therefore one is able to rewrite them as:
\begin{subequations} \label{schw_metrics_2}
\begin{equation}
\mathrm{d}s^2_{in}=-\left(\frac{2M}{t}-1\right)^{-1}\mathrm{d}t^2+\left(\frac{2M}{t}-1\right)\mathrm{d}r^2+t^2\left(\mathrm{d}\theta^2+\sin^2\theta\,\mathrm{d}\phi^2\right),
\end{equation}
\begin{equation}
\mathrm{d}s^2_{out}=-\left(\frac{2(M+m)}{t}-1\right)^{-1}\mathrm{d}t^2+\left(\frac{2(M+m)}{t}-1\right)\mathrm{d}r^2+t^2\left(\mathrm{d}\theta^2+\sin^2\theta\,\mathrm{d}\phi^2\right),
\end{equation} \label{Schw_metrics}
\end{subequations}
where all the metric coefficients are positive.

This allows for mapping of each metric into a KS-type metric which, in the Misner parameterization, is given by:
\begin{equation}
\mathrm{d}s^2=-N^2(t)\mathrm{d}t^2+e^{2\sqrt{3}\beta(t)}\mathrm{d}r^2+e^{-2\sqrt{3}(\beta(t)+\Omega(t))}\left(\mathrm{d}\theta^2+\sin^2\theta\,\mathrm{d}\phi^2\right).
\end{equation}
This is accomplished by using the map given in Ref. \cite{Catarina_2010},
\begin{equation} \label{map}
N^2=\left(\frac{2M}{t}-1\right)^{-1}, \quad e^{2\sqrt{3}\beta}=\left(\frac{2M}{t}-1\right), \quad e^{-2\sqrt{3}(\beta+\Omega)}=t^2,
\end{equation}
so that the two metrics can be written as:
\begin{subequations} \label{initialKSmetrics}
\begin{equation}
\mathrm{d}s^2_{in}=-N_{+}^2\mathrm{d}t^2+e^{2\sqrt{3}\beta_+}\mathrm{d}r^2+e^{-2\sqrt{3}(\beta_++\Omega_+)}\left(\mathrm{d}\theta^2+\sin^2\theta\,\mathrm{d}\phi^2\right),
\end{equation}
\begin{equation}
\mathrm{d}s^2_{out}=-N_{-}^2\mathrm{d}t^2+e^{2\sqrt{3}\beta_-}\mathrm{d}r^2+e^{-2\sqrt{3}(\beta_-+\Omega_-)}\left(\mathrm{d}\theta^2+\sin^2\theta\,\mathrm{d}\phi^2\right),
\end{equation}
\end{subequations}
where $N_\pm$, $\beta_\pm$ and $\Omega_\pm$ are functions of the $t$ coordinate alone. Using this procedure one is able to carry out the quantization of the above stablished geometrical system maintaining all the symmetries of the problem. Nevertheless, before proceeding, we must be aware of some subtleties of this map. First, the relationship between $\Omega$ and $t$ is not injective, being symmetric around the point $t=M$, the value for which $\Omega(t)$ has a minimum. This brings up no condition on the above parameterization, however, as the the map should return into a single set $(\beta,\Omega)$ for any given value of $t$, and since the function $\beta(t)$ is injective, the parameterization is indeed fiducial. Moreover, given the aforementioned symmetry for $\Omega(t)$, the distinction between each point symmetrically distanced from $t=M$ is made by the $\beta(t)$ function, having positive values for $t<M$ and negative values for $t>M$. A final remark concerning this map is related to the minimun of $\Omega(t)$. This point is reached at $\sim -\ln(M^2)$ which is negative for $M > 1$. However, because this minimun is finite, one can, without loss of generality, consider $M\leq 1$ and analyze the results for $\Omega>0$. These properties of the map $t\rightarrow (\beta,\Omega)$ will be relevant in the analysis of the resulting wavefunction in Sec.~\ref{quantization_section}.

\section{The Action} \label{action_section}

The first step in the quantization of a geometrical system is to write its action. Since the system consists of two different spacetimes, in contact at a boundary defined by the shell, two difficulties arise: first, there is a boundary in the system (an external boundary for the inner spacetime and an internal boundary for the outer spacetime), which requires the inclusion of a Gibbons-Hawking-York like term into the action; second, there is a discontinuity in the Ricci curvature scalar, which does affect the Einstein-Hilbert part of the action \cite{Barrabes_2003,Poisson_2004}. Therefore, the resulting action is of the form:
\begin{equation} \label{fullaction_1}
S=S_{\mathrm{EH}}+S_{\mathrm{boundary}}=\frac{1}{16\pi}\int\sqrt{-g}\, \mathrm{d}^4x\Big[R_-\Theta(u-u_0)+R_+\Theta(-u+u_0)+\delta R\Big]+S_{\mathrm{boundary}},
\end{equation}
where natural units, $c=G=\hbar=1$, have been considered and $u$ is a coordinate that defines the position of the shell, a hypersurface $\Sigma$ in the embedding manifold $\mathcal{M}=\mathcal{M}^+\cup\mathcal{M}^-\cup\Sigma$. The term $\delta R$ is the result of the discontinuity in quantities computed on both sides of the shell, and the boundary term, $S_{\mathrm{boundary}}$, shall include two contributions, since the shell acts as a boundary for the inner and outer spacetimes, with quantities defined at each boundary depending on the metric of the spacetime.
\par
It is useful to introduce a change of coordinates in the metrics given by Eq.~(\ref{initialKSmetrics}) $u=r-\zeta t^*$, where $\zeta=\pm1$ and $\mathrm{d}t^*\defeq N e^{-\sqrt{3}\beta}\mathrm{d}t$. The two values of $\zeta$ correspond to either a collapsing shell ($\zeta=-1$) or an expanding one ($\zeta=1$). The metrics then take the form:
\begin{equation} \label{metric_u_coords}
\mathrm{d}s^2=e^{\sqrt{3}\beta}\,\mathrm{d}u\left(e^{\sqrt{3}\beta}\,\mathrm{d}u+2\zeta N\,\mathrm{d}t\right)+e^{-2\sqrt{3}\Lambda}\,\left(\mathrm{d}\theta^2+\sin^2\theta\,\mathrm{d}\phi^2\right),
\end{equation} 
where the $\pm$ symbols have been dropped and the combination $\Lambda=\beta+\Omega$ is introduced. Of course, the two metrics in Eq.~(\ref{initialKSmetrics}) are expressed in the form of Eq.~(\ref{metric_u_coords}). The system depends now on the three functions $N$, $\beta$ and $\Lambda$, on each side of the shell. Also, in these coordinates, the shell is located at a surface of constant $u$, i.e. $u=u_0$, which simplifies the forthcoming analysis.

\subsection{The discontinuity term - $\delta R$}
In order to further proceed, we must endow the space with a more robust geometric structure. For this, a basis of vectors for each side of the shell is constructed. A particularly useful choice is the set composed by the normal vector, the holonomic vectors on $\Sigma$ for $\theta$ and $\phi$, and a fourth vector to be introduced shortly. The normal vectors to a null hypersurface are given by:
\begin{equation}
n_\pm^\mu=\chi_\pm^{-1}(t,u)\,g^{\mu\nu}_\pm\partial_\nu\Phi(u),
\end{equation}
where $\Phi(u)=0$ defines $\Sigma$, and $\chi_\pm$ is a normalization factor that can be chosen arbitrarily. Thus, in this scenario $\Phi(u)=u-u_0$, and so:
\begin{equation}
n_\pm^\mu=-\frac{\zeta  e^{-\sqrt{3} \beta _\pm}}{N_{\pm} \chi _\pm}\partial_t.
\end{equation}
From here on, it is assumed that the time coordinate in both sides of the hypersurface is the same, so that the whole manifold can be parameterized by the same temporal parameter\footnote{For a more general setup with different time coordinates, see the Appendix.}. With this choice, a natural set of coordinates on $\Sigma$ is $(t,\theta,\phi)$ since those coincide if the two sides are connected at a constant value $u=u_{0}$. Thus, the holonomic vectors are given by:
\begin{equation}
e^\mu_{(A)}\defeq \frac{\partial x^\mu}{\partial y^A}=\delta^\mu_A\partial_A,
\end{equation}
for $A=\theta,\phi$ and by
\begin{equation}
e^\mu_{(t)}\defeq \frac{\partial x^\mu}{\partial t}=-\partial_t,
\end{equation}
for $t$.
With these vectors, we can define the induced metric on the hypersurface $\Sigma$ as $g_{ab}=e^\mu_{(a)}e^\nu_{(b)}g_{\mu\nu}$, which yields:
\begin{equation}
\mathrm{d}s^2_{|\Sigma_{\pm}}=e^{-2\sqrt{3}\Lambda_\pm}\left(\mathrm{d}\theta^2+\sin^2\theta\,\mathrm{d}\phi^2\right).
\end{equation}
For this hypersurface to possess a well-defined geometry, the induced metric defined on it must be unique. This allows for imposing the first condition on the metric fields as $\Lambda_{|\Sigma-}=\Lambda_{|\Sigma+}$. However, since $\Lambda_\pm$ only depends on the coordinate $t$ and not on $u$, the junction condition is valid for any $t$ and therefore for the whole manifold. Moreover, the coordinate $t$ covers the entire manifold, thus $\dot{\Lambda}_+|_{u=u_0}=\dot{\Lambda}_-|_{u=u_0}$.

In order to complete the basis of vector fields, an auxiliary one must be introduced (one at each side of $\Sigma$), because $n^\mu\propto e^\mu_{(t)}$. This additional vector field, $M^\mu$, must not be orthogonal to $n^\mu$, otherwise it would be a vector on $\Sigma$. Additionally, it shall be assumed null and orthogonal to the other basis vectors, i.e.
\begin{equation}
M^\mu\,M_\mu=0, \qquad M\cdot n=\varepsilon^{-1}\neq 0, \quad M\cdot e_{(A)}=0,
\end{equation}
for $A=\theta,\phi$ and $\varepsilon$ arbitrary. The vector is thus determined (up to the constant factor) as:
\begin{equation}
M^\mu_\pm=\frac{\chi _\pm(u,t)}{\varepsilon _\pm}\partial_u-\frac{\zeta  e^{\sqrt{3} \beta _\pm} \chi _\pm(u,t)}{2 \varepsilon _\pm N_\pm}\partial_t.
\end{equation}
From now on, a complete basis of vector fields for both sides of the hypersurface $\Sigma$ is established, namely $\left(n^\mu,M^\mu,e_{(\theta)}^\mu,e_{(\phi)}^\mu\right)$. It allows for defining the quantities intrinsic to $\Sigma$, which do not depend on their four-dimensional counterparts (although they may be related in some way). These are the intrinsic quantities that will be used to solve the problem. For this purpose, the four-vectors are projected into the basis onto  $\Sigma$. The normal vector can be written as $n^\mu=n^a\,e^\mu_{(a)}$, where its components are:
\begin{equation}
n^a_\pm=\frac{\zeta  e^{-\sqrt{3} \beta _\pm} }{N_\pm \chi _\pm(u,t)}\partial_t.
\end{equation}
The distinction between the four-vector and the three-vector is made through the indices: greek indices represent four-dimensional quantities, lowercase latin indices represent three-dimensional quantities and uppercase latin indices represent two-dimensional quantities, particularly $\theta$ and $\phi$. For the transverse vector, the one-form $M_\mu$ is projected onto $\Sigma$ to give:
\begin{equation}
M_a^\pm=M_\mu^\pm e^\mu_{(a)}=\frac{\zeta  N_\pm e^{\sqrt{3} \beta _\pm} \chi _\pm(u,t)}{ \varepsilon _\pm }\partial_t.
\end{equation}
The requirement that the quantities $n^a$ and $M_a$ coincide on both sides of the shell determines a unique structure for $\Sigma$. This condition imposes restrictions on the normalization factors, that is:
\begin{equation} \label{boundaryCondition}
    \frac{   e^{-\sqrt{3} \beta _-}}{N_- \chi _-}\Big|_{u=u_0}=\frac{   e^{-\sqrt{3} \beta _+} }{N_+ \chi_+}\Big|_{u=u_0},
\end{equation}
and,
\begin{equation}
    \frac{  N_- e^{\sqrt{3} \beta _-} \chi _-}{ \varepsilon _- }\Big|_{u=u_0}=\frac{  N_+ e^{\sqrt{3} \beta _+} \chi _+}{ \varepsilon _+ }\Big|_{u=u_0}.
\end{equation}
These two equations together require that $\varepsilon_-=\varepsilon_+$. This is a sensible result since $\varepsilon=M\cdot n=M_\mu n^\mu=M_a n^a$. Using the condition Eq.~(\ref{boundaryCondition}) throughout the calculations, it is possible to verify that the resulting term for $\delta R$ does not depend on the chosen normalization.
\par
Having defined all the quantities necessary to study the structure of the hypersurface as a submanifold, one proceeds to compute the transverse curvature, defined by \cite{Barrabes_2003}:
\begin{equation}
\mathcal{K}_{ab}=M_\mu\,e^\mu_{(a);\nu}\,e^\nu_{(b)},
\end{equation}
where the semicolon denotes covariant derivatives. This tensor quantity is used instead of the more common extrinsic curvature ($K_{ab}$) as it relies on the changes of the metric along its normal vector which, in the case of null surfaces, is not a quite suitable choice. This occurs since the normal vector is also part of the hypersurface and does not carry information about the embedding space, resulting in an extrinsic curvature that is continuous across null surfaces (and across $\Sigma$ in particular). Hence, the transverse curvature, which relies on the change of the metric along the introduced transverse vector, is used instead \cite{Barrabes_2003}. This can be computed for both sides of the hypersurface, for which the non-vanishing components are:
\begin{equation}
\begin{aligned}
\mathcal{K}_{tt}^\pm&=-\frac{\zeta  e^{\sqrt{3} \beta _\pm} \chi _\pm}{ \varepsilon} \left(\dot{N}_\pm +\sqrt{3}N_\pm \dot{\beta} _\pm\right), \\
\mathcal{K}_{\theta\theta}^\pm&=\frac{\sqrt{3} \zeta  e^{\sqrt{3} \left(\beta _\pm-2 \Lambda _\pm\right)} \dot{\Lambda} _\pm \chi _\pm}{2 \varepsilon N_\pm}, \\
\mathcal{K}_{\phi\phi}^\pm&=\mathcal{K}_{\theta\theta}^\pm\sin ^2(\theta ).
\end{aligned}
\end{equation}
The transverse curvature is a tensor defined only on $\Sigma$ and it can be used to extract the energy-momentum tensor on this surface. For this, a new tensor, $\gamma_{ab}$, is defined in the following way:
\begin{equation}
\gamma_{ab}\defeq 2\left[\mathcal{K}_{ab}\right],
\end{equation}
also only on $\Sigma$. Additionally, in order to compute the discontinuity in the energy-momentum tensor, it is necessary to find an equivalent to an inverse induced metric. This cannot be done in the usual way, since the induced metric is degenerate. However, it is possible to construct a ``pseudo-inverse'' tensor, $g_*^{ab}$, which obeys \cite{Barrabes_2003}:
\begin{equation}
g^{\mu\nu}=g_*^{ab}\,e^\mu_{(a)}\,e^\nu_{(b)} + \varepsilon\,n^a\,(e_{(a)}^{\mu}N^{\nu}+e_{(a)}^{\nu}N^{\mu}).
\end{equation}
For the present problem, this three dimensional tensor can be written as:
\begin{equation}
g_*^{ab}=\left(
\begin{array}{ccc}
 0 & 0 & 0 \\
 0 & e^{2 \sqrt{3} \Lambda} & 0 \\
 0 & 0 & e^{2 \sqrt{3} \Lambda } \csc ^2(\theta ) \\
\end{array}
\right),
\end{equation}
where each non-vanishing entry is the inverse of the corresponding entry in the induced metric. With this object, the surface energy-momentum tensor is computed to be:
\begin{equation}
T^{ab}_{|_\Sigma}=\frac{1}{16\pi}\Big[-\left(\gamma_{cd}\,g_*^{cd}\right)n^an^b-\left(\gamma_{cd}\,n^cn^d\right)g_*^{ab}+\left(g_*^{ac}\,n^bn^d+g_*^{bc}\,n^an^d\right)\gamma_{cd}\Big],
\end{equation}
for which its non-vanishing components are:
\begin{equation}
\begin{aligned}
T^{tt}_{|_\Sigma}&=\frac{\sqrt{3} \zeta}{8 \pi\varepsilon }  \left[\frac{e^{-\sqrt{3} \beta _-} \dot{\Lambda} _-}{N_-^3 \chi _-}-\frac{e^{-\sqrt{3} \beta _+} \dot{\Lambda} _+}{ N_+^3 \chi _+}\right], \\
T^{\theta\theta}_{|_\Sigma}&=-\frac{\zeta e^{- \sqrt{3} \left(\beta _--2\Lambda _-\right)} \left(\dot{N}_- +\sqrt{3}N_-  \dot{\beta} _- \right)}{8 \pi  \varepsilon N_-^2 \chi _-}+\frac{\zeta e^{- \sqrt{3} \left(\beta _+-2\Lambda _+\right)} \left(\dot{N}_+ \sqrt{3}+N_+ \dot{\beta} _+\right)}{8 \pi  \varepsilon N_+^2 \chi _+}, \\
T^{\phi\phi}_{|_\Sigma}&=\frac{T^{\theta\theta}}{\sin^2(\theta)}.
\end{aligned}
\end{equation}
Since this tensor is defined on $\Sigma$, we can compute its trace using the induced metric on this hypersurface, which yields:
\begin{equation}
\begin{aligned}
T_{|_\Sigma}=-\frac{\zeta  e^{-\sqrt{3} \beta _-} \left(\dot{N}_- +\sqrt{3}\dot{N}_-  \dot{\beta} _- \right)}{4 \pi  \varepsilon  N_-^2 \chi _-}+\frac{\zeta  e^{-\sqrt{3} \beta _+} \left(\dot{N}_+ +\sqrt{3}\dot{N}_+  \dot{\beta} _+ \right)}{4 \pi  \varepsilon  N_+^2  \chi _+}.
\end{aligned}
\end{equation}
The construction so far, as well as the requirements on the quantities on both sides, have been based on the assumption that the geometrical setup is a solution of the Einstein field equations (EFE). In particular, the definition of the surface stress-energy tensor follows from the application of these equations as it can be seen in Refs. \cite{Barrabes_2003,Poisson_2004}. Therefore, the surface scalar curvature can be obtained directly from the trace of the stress-energy tensor by taking the trace of the EFE. This leads to:
\begin{equation}
R_{|_\Sigma}=-8\pi T_{|_\Sigma},
\end{equation}
which, for the particular geometry at hand, yields:
\begin{equation}
\begin{aligned}
R_{|_\Sigma}=\frac{2\zeta  e^{-\sqrt{3} \beta _-} \left(\dot{N}_- +\sqrt{3}\dot{N}_-  \dot{\beta} _- \right)}{  \varepsilon  N_-^2 \chi _-}-\frac{2\zeta  e^{-\sqrt{3} \beta _+} \left(\dot{N}_+ +\sqrt{3}\dot{N}_+  \dot{\beta} _+ \right)}{ \varepsilon  N_+^2  \chi _+},
\end{aligned}
\end{equation}
where Eq.~(\ref{boundaryCondition}) has been used in order to set the resulting curvature scalar in a more symmetric fashion. The discontinuity in the scalar curvature, $\delta R$, is then obtained as \cite{Barrabes_2003}:
\begin{equation}
\delta R=\varepsilon\,R_{|_\Sigma}\chi(u,r)\delta(u-u_0),   
\end{equation}
resulting in the expression:
\begin{equation}
\begin{aligned}
\delta R=\left(\frac{2\zeta  e^{-\sqrt{3} \beta _-} \left(\dot{N}_- +\sqrt{3}\dot{N}_-  \dot{\beta} _- \right)}{   N_-^2 }-\frac{2\zeta  e^{-\sqrt{3} \beta _+} \left(\dot{N}_+ +\sqrt{3}\dot{N}_+  \dot{\beta} _+ \right)}{  N_+^2}\right)\,\delta(u-u_0),
\end{aligned}
\end{equation}
which, in fact, is independent of the choice of $\chi$ and $\eta$. We are now able to introduce this expression into the action and, after integrating over $\theta$ and $\phi$, the resulting action for $\Sigma$ coming from the Einstein-Hilbert (EH) action yields:
\begin{equation} \label{term1}
\begin{aligned}
S_{\mathrm{EH}}^\Sigma=\frac{1}{2}\int_{\Sigma} \left[\frac{2\zeta  e^{-2\sqrt{3} \Lambda _-} \left(\dot{N}_- +\sqrt{3}\dot{N}_-  \dot{\beta} _- \right)}{   N_- }-\frac{2\zeta  e^{-2\sqrt{3} \Lambda _+} \left(\dot{N}_+ +\sqrt{3}\dot{N}_+  \dot{\beta} _+ \right)}{  N_+}\right]\mathrm{d}t.
\end{aligned}
\end{equation} 

\subsection{The Boundary term}

Additionally to the surface term given by Eq.~(\ref{term1}) as discussed in the previous section, an action term related to the boundary of the manifolds must be included. 
For spacelike or timelike boundaries this is expressed in terms of the Hawking-Gibbons-York term, which is given by:
\begin{equation}
S_{\mathrm{GHY}}=\frac{\epsilon}{8\pi}\int_{\partial \mathcal{V}}\mathrm{d}^3y\,\sqrt{h}K,
\end{equation}
where $K$ is the trace of the extrinsic curvature, $\epsilon=+1$ $(-1)$ if $\partial \mathcal{V}$ is timelike (spacelike) and the $y$ integration is over the coordinates at the boundary of the manifold. In the present problem, however, the boundary of the manifold is null-like, and thus this term is not well-defined, since, for example, the induced metric is singular and hence its determinant vanishes. In a recent work, an analogue term for null surfaces was proposed \cite{Parattu_2016} and in another it is extended for any boundary comprising any number of timelike, spacelike and null boundary segments \cite{Lehner_2016}. In the present scenario, the boundary is null, and so one may consider the approach of Ref. \cite{Parattu_2016}:
\begin{equation} \label{nullboundaryaction}
S_{\mathrm{boundary}}=\frac{1}{8\pi}\int_{\partial \mathcal{V}}\mathrm{d}y_A^3\sqrt{-g}\chi_{\pm}(\Theta+\kappa),
\end{equation}
where $y_A$ are the coordinates on $\Sigma$, $\kappa$ is the parameter that measures the ``non-affinity'' of the parameter on the null generators, $\lambda=t$, and $\Theta$ is defined as:
\begin{equation}
\Theta=q^{\mu\nu}\Theta_{\mu\nu}, \quad \Theta_{\mu\nu}=q^\sigma_\mu q^\rho_\nu n_{\rho;\sigma},
\end{equation}
where $q_{\mu\nu}$ is the induced two dimensional metric on $\Sigma$, extended to four dimensions. This metric can be computed as:
\begin{equation}
q_{AB}=g^{ab}\,e^a_{(A)}\,e^b_{(B)}=\left(
\begin{array}{cc}
 e^{-2 \sqrt{3} \Lambda _\pm(t)} & 0 \\
 0 & e^{-2 \sqrt{3} \Lambda _\pm(t)} \sin ^2(\theta ) \\
\end{array}
\right) , \quad {A,\,B}\,=\theta,\phi.
\end{equation}
It is simple to obtain the inverse, $q^{AB}$, as a regular inverse matrix. Then, the extension to four dimensions is obtained by using the basis vectors:
\begin{equation}
q^{\mu\nu}=q^{AB}\,e^\mu_{(A)}\,e^\nu_{(B)}=\left(
\begin{array}{cccc}
 0 & 0 & 0 & 0 \\
 0 & 0 & 0 & 0 \\
 0 & 0 & e^{2 \sqrt{3} \Lambda _\pm(t)} & 0 \\
 0 & 0 & 0 & e^{2 \sqrt{3} \Lambda _\pm(t)} \csc ^2(\theta ) \\
\end{array}
\right).
\end{equation}
These are uniquely defined on $\Sigma$ since $\left[\Lambda\right]=0$. 
\par
The computation of $\Theta_{\mu\nu}$ is rather straightforward. However, one must compute two tensors, one for the border of the inner space and the other for the outer space. The non-vanishing terms of this tensor are then:
\begin{equation}
\begin{aligned}
\Theta_{\theta\theta}^\pm&=-\frac{\sqrt{3} \zeta  e^{-\sqrt{3} \left(\beta _\pm+2 \Lambda _\pm\right)} \dot{\Lambda} _\pm}{N_\pm \chi _\pm}, \\
\Theta_{\phi\phi}^\pm&=\Theta_{\theta\theta}^\pm \sin^2(\theta).
\end{aligned}
\end{equation}
It follows that its trace is given by:
\begin{equation}
\Theta_\pm=-\frac{2 \sqrt{3} \zeta  e^{-\sqrt{3} \beta _\pm} \dot{\Lambda} _\pm}{N_\pm \chi _\pm}.
\end{equation}
Considering this result and the boundary action given in Eq.~(\ref{nullboundaryaction}), it is straightfrorward to see that this term does not depend on the specific normalization chosen for the normal vectors to the hypersurface.
\par
Regarding $\kappa$, it can be computed by evaluating if the parameter on the null generators is affine. This yields:
\begin{equation}
n^\mu n^\nu_{;\mu}=\kappa n^\mu\Leftrightarrow\left(\frac{\zeta  e^{-\sqrt{3} \beta _\pm} \dot{\chi} _\pm}{N_\pm \chi _\pm{}^3},0,0,0\right)=\kappa\left(-\frac{1}{\chi _+},0,0,0\right),
\end{equation}
thus $\kappa$ is found to be:
\begin{equation}
\kappa=-\frac{\zeta  e^{-\sqrt{3} \beta _\pm} \dot{\chi} _\pm}{N_\pm \chi _\pm{}^2}.
\end{equation}
The first observation regarding this term is that, when substituted into Eq.~(\ref{nullboundaryaction}), it seems to depend on the normalization chosen for the null vectors. However, this is not true once the boundary conditions are imposed.

\subsection{Full boundary action}

When all the factors that contribute to the boundary part of the action are taken into account, we can write:
\begin{equation}
S_{\mathrm{boundary}}=\frac{1}{2}\int \mathrm{d}t\left[\sqrt{-g_+}\chi_+\left(\varepsilon\,\delta R_++2\Theta_++2\kappa_+\right)-\sqrt{-g_-}\chi_-\left(\varepsilon\,\delta R_-+2\Theta_-+2\kappa_-\right)\right].
\end{equation}
We now apply the boundary condition, Eq.~(\ref{boundaryCondition}), to the above action. It turns out that, not only the dependence on the derivative of $\chi_\pm$ cancels out, but also that the whole boundary action vanishes. This implies that, at least in the particular setup used in this work, the boundary terms that arise from the discontinuity in the derivatives of the metric fields are cancelled out by the terms which must be added to the Einstein-Hilbert action in the case of null boundaries. As a result, only the bulk terms of the action are relevant for the problem.

We can wonder why this precise cancelation happens since it appears to imply that the two spacetime pieces, each with its own metric, do not influence on each other and that the analysis could be carried for each piece separately. First, it should be clear that the cancelation happens because a null boundary was inserted in each of the spacetimes and hence those boundaries were matched while maintaining the full spacetime as a solution of EFE. As will be seen in the upcoming treatment, the existence of this matching will lead to an Hamiltonian which is different from the one that would be obtained from a KS manifold alone. Second, it is natural that this separation occurs, solely based on the behavior on null geodesics. These geodesics are determined completely by their null nature, unlike timelike ones, which depend on the conserved quantities of the underlying spacetime. Therefore, for an observer inside (outside) the shell, the trajectory it follows is completely determined by the requirement that it must be null on that spacetime and does not depend on anything else. This implies that trajectories are determined apart from the boundary conditions, hence the separation. Nevertheless, this does not mean that it is possible to connect whatever geometry by mathcing null surfaces with each other, since consistent matching conditions cannot be found in general.

\subsection{Full action and Hamiltonian}
Now that the action for the boundary terms has been determined and proven to vanish in our scenario, we need only to determine the bulk terms of the action. Since the spacetimes have the same structure, i.e., the metric components are written in the same way, the form of the Lagrangian for the inner and outter spaces are the same. Taking a metric of the form Eq.~(\ref{metric_u_coords}), the bulk action is straightforward to compute:
\begin{equation}
S_{\mathrm{bulk}}=\frac{1}{2}\int \mathrm{d}t\,\mathrm{d}u \left(e^{\sqrt{3}\beta}N+\frac{e^{-\sqrt{3}(\beta+2\Omega)}}{N}(3\dot{\beta}^2-3\dot{\Omega}^2)\right).
\end{equation}
This leads to the Hamiltonian obtained in Refs. \cite{Catarina_2008} (a slight rescaling of the $\Omega$ variable leads to a different coefficient on the potential term). Since all the terms have been computed, they can be inserted into Eq.~(\ref{fullaction_1}) resulting in the expression:
\begin{equation}
\begin{aligned}
S=&\frac{1}{2}\int \mathrm{d}t\,\mathrm{d}u\left[\left(e^{\sqrt{3}\beta_-}N_-+\frac{e^{-\sqrt{3}(\beta_-+2\Omega_-)}}{N}(3\dot{\beta}_-^2-3\dot{\Omega}_-^2)\right)\Theta(u-u_0)+ \right. \\
&\left. +\left(e^{\sqrt{3}\beta_+}N_++\frac{e^{-\sqrt{3}(\beta_++2\Omega_+)}}{N}(3\dot{\beta}_+^2-3\dot{\Omega}_+^2)\right)\Theta(u_0-u)\right].
\end{aligned}
\end{equation}
In order to find the Hamiltonian associated with this action, one first imposes the suitable boundary conditions, choosing the normalization factors as functions of $t$ alone, i.e. $\chi_\pm(t)$. This is still consistent with the matching conditions on $\Sigma$ and in fact, this choice extends their validity to the whole manifold. Thus the action can be written as:
\begin{equation}
\begin{aligned}
S=&\int \mathrm{d}t\,\mathrm{d}u\left[\frac{1}{2}\,e^{\sqrt{3}\beta_-}N_-\left(\frac{\chi_-}{\chi_+}\Theta(u_0-u)+\Theta(u-u_0)\right)+ \right. \\
&\left. +\frac{3}{2\,N_-}e^{-\sqrt{3}\beta_-}\left(\frac{\chi_+}{\chi_-}e^{-2\sqrt{3}\Omega_+}\left(\dot{\beta}_+^2-\dot{\Omega}_+^2\right)\Theta(u_0-u)+e^{-2\sqrt{3}\Omega_-}\left(\dot{\beta}_-^2-\dot{\Omega}_-^2\right)\Theta(u-u_0)\right)\right].
\end{aligned}
\end{equation}
We can now determine the momenta associated with this Lagrangian as:
\begin{subequations}
\begin{equation}
p_{\beta_-}=\frac{3 e^{-\sqrt{3} \left(\beta _-+2 \Omega _-\right)} \dot{\beta} _-}{N_-}\Theta (-u+u_0),
\end{equation}
\begin{equation}
p_{\beta_+}=\frac{3 e^{-\sqrt{3} \left(\beta _-+2 \Omega _+\right)} \dot{\beta} _+}{N_+}\frac{\chi_+}{\chi_-}\Theta (u-u_0),
\end{equation}
\begin{equation}
p_{\Omega_-}=-\frac{3 e^{-\sqrt{3} \left(\beta _-+2 \Omega _-\right)} \dot{\Omega} _-}{N_-}\Theta (-u+u_0),
\end{equation}
\begin{equation}
p_{\Omega_+}=-\frac{3 e^{-\sqrt{3} \left(\beta _-+2 \Omega _+\right)} \dot{\Omega} _+}{N_+}\frac{\chi_+}{\chi_-}\Theta (u-u_0),
\end{equation}
\end{subequations}
where for the Heaviside distribution, $\Theta(u-u_0)^2=\Theta(u-u_0)$. One possible route to determine the Hamiltonian constaint of the system is to demand that the variation of the action with respect to the lapse function vanishes. Doing this, we get the Hamiltonian given by:
\begin{equation}
H=e^{2 \sqrt{3} \Omega _-}\left(p_{\beta_-}^2-p_{\Omega_-}^2\right)+e^{2 \sqrt{3} \Omega _+}\frac{\chi_-}{\chi_+}\left(p_{\beta_+}^2-p_{\Omega_+}^2\right)-3\left(\frac{\chi_-}{\chi_+}\Theta(u_0-u)+\Theta(u-u_0)\right).
\end{equation}
There is still the freedom to choose the ratio $\chi_-/\chi_+$ in Eq. (\ref{boundaryCondition}); a particularly useful choice is $\chi_-/\chi_+=1$. Under this choice the Hamiltonian becomes:
\begin{equation}
H=e^{2 \sqrt{3} \Omega _-}\left(p_{\beta_-}^2-p_{\Omega_-}^2\right)+e^{2 \sqrt{3} \Omega _+}\left(p_{\beta_+}^2-p_{\Omega_+}^2\right),
\end{equation}
where a constant term is absorbed into the definition of the Hamiltonian. This Hamiltonian is similar to the one for the KS spacetime, but not quite the same. The difference lies in the matching conditions for the null boundaries which allows for eliminating the constant term.

\section{Quantization} \label{quantization_section}
\subsection{Commutative scenario}
\begin{figure}[h]
    \includegraphics[width=0.8\textwidth]{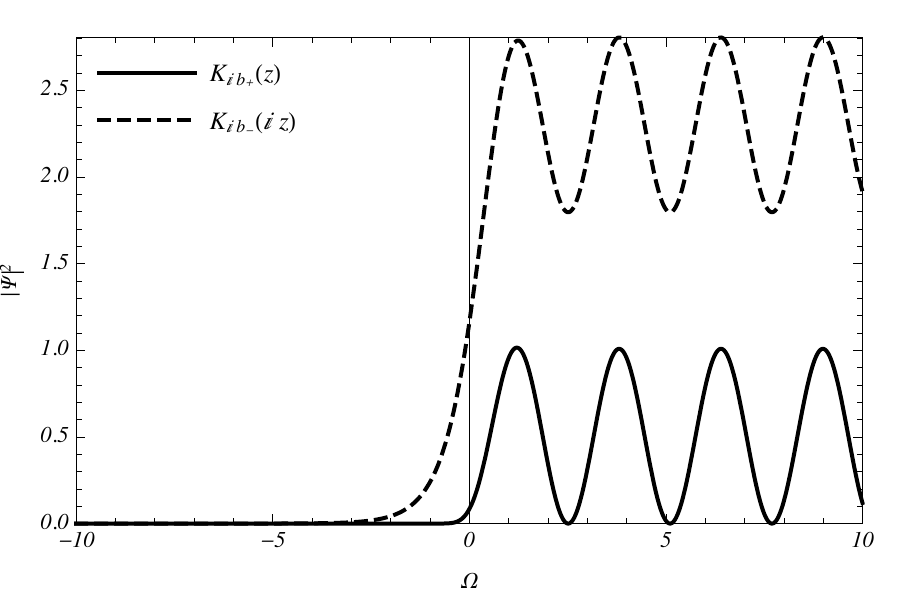}
    \caption{$\left|\Psi_{-}\right|^2$ and $\left|\Psi_{+}\right|^2$ as function of $\Omega_\pm$. The parameters are $b_\pm = 0.7$ and $\lambda=2$.}
    \label{Psi_plus_and_minus}
\end{figure}

\begin{figure}[h]
    \includegraphics[width=0.8\textwidth]{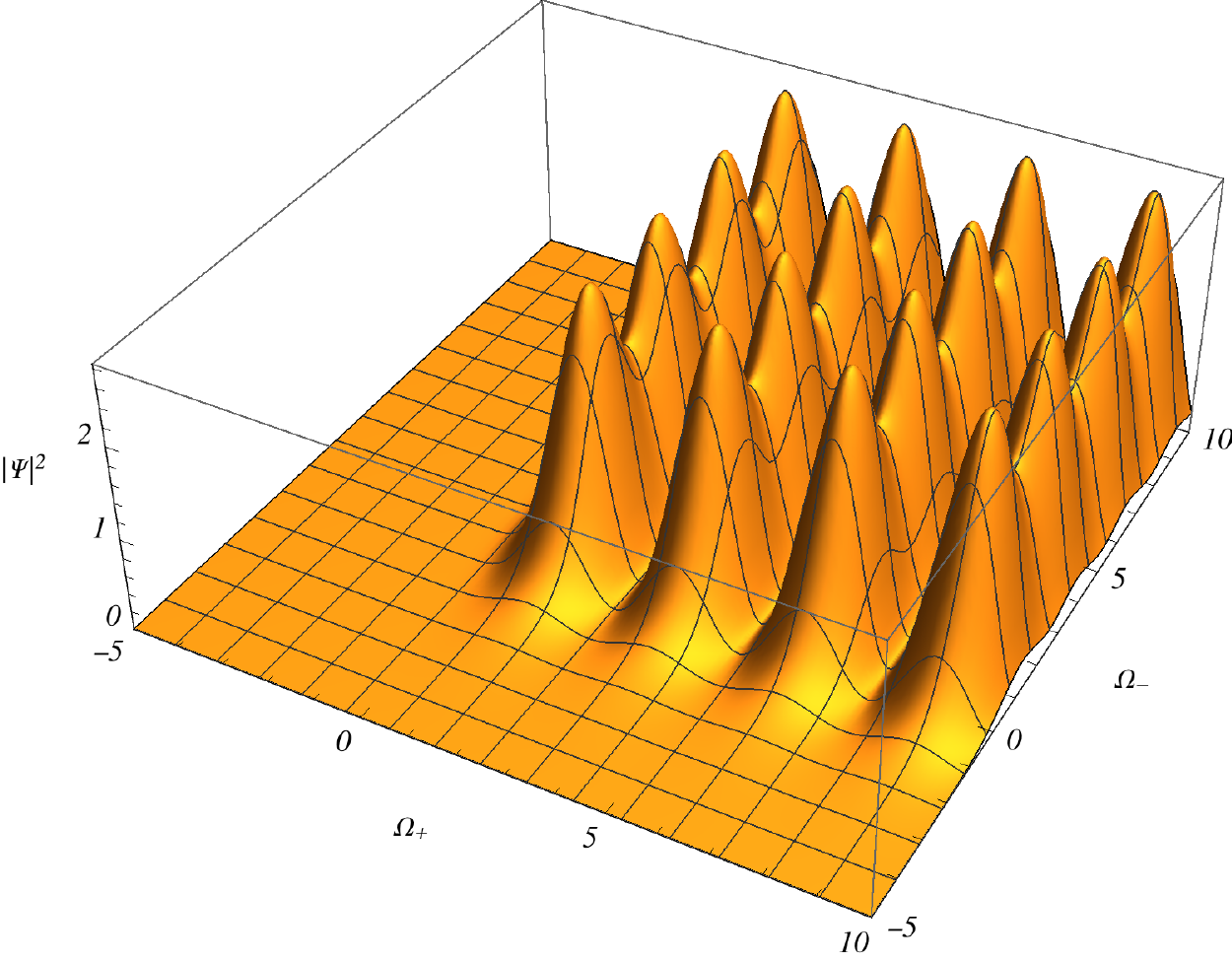}
    \caption{$\left|\Psi\right|^2$ in the coordinate space of $\Omega_\pm$. The parameters are $b_\pm = 0.7$ and $\lambda^2=2$.}.
    \label{Psi_total}
\end{figure}

For the quantum Hamiltonian operator written as:
\begin{equation}
\hat{H}=e^{2 \sqrt{3} \hat{\Omega} _-}\left(\hat{p}_{\beta_-}^2-\hat{p}_{\Omega_-}^2\right)+e^{2 \sqrt{3} \Omega _+}\left(\hat{p}_{\beta_+}^2-\hat{p}_{\Omega_+}^2\right),
\end{equation}
the canonical quantization method, once applied to the Hamiltonian constraint, results into the Wheeler-deWitt equation for the problem:
\begin{equation}
\left[-e^{2 \sqrt{3} \Omega _-}(\partial^2_{\beta_-}-\partial_{\Omega_-}^2)-e^{2 \sqrt{3} \Omega _+}(\partial^2_{\beta_+}-\partial_{\Omega_+}^2)\right]\Psi(\beta_-,\Omega_-,\beta_+,\Omega_+)=0,
\label{wdee}
\end{equation}
where the Heisenberg-Weyl operator representation and ordering have been adopted.
Eq.~\eqref{wdee} shows no crossing terms between quantities of inner and outer manifolds. This suggests that the wave function for the problem is separable and can be written as $\Psi(\beta_-,\Omega_-,\beta_+,\Omega_+)=\Psi_-(\beta_-,\Omega_-)\Psi_+(\beta_+,\Omega_+)$, which leads to the following decoupled equations:
\begin{subequations}
\begin{equation}
-e^{2 \sqrt{3} \Omega _-}(\partial^2_{\beta_-}-\partial_{\Omega_-}^2)\Psi_-(\beta_-,\Omega_-)=\lambda^2\Psi_-(\beta_-,\Omega_-),
\end{equation}
\begin{equation}
-e^{2 \sqrt{3} \Omega _+}(\partial^2_{\beta_+}-\partial_{\Omega_+}^2)\Psi_+(\beta_+,\Omega_+)=-\lambda^2\Psi_+(\beta_+,\Omega_+),
\end{equation}
\end{subequations}
which describe a dynamics similar to that one of the KS cosmological problem \cite{Catarina_2008,Compean_2002}, with a real eigenvalue, $\lambda$. The solutions for $\Psi_+$ and $\Psi_-$ are given by:
\begin{subequations} \label{Commutative_sols}
\begin{equation}
\Psi_{-}(\beta_-,\Omega_-)=e^{\mathrm{i}\sqrt{3}b_-\beta_-}K_{\mathrm{i}b_-}\left(\frac{\lambda}{\sqrt{3}}e^{-\sqrt{3}\Omega_-}\right),
\end{equation}
\begin{equation}
\Psi_{+}(\beta_+,\Omega_+)=e^{\mathrm{i}\sqrt{3}b_+\beta_+}K_{\mathrm{i}b_+}\left(\frac{\mathrm{i}\lambda}{\sqrt{3}}e^{-\sqrt{3}\Omega_+}\right),
\end{equation}
\end{subequations}
which results into the full solution:
\begin{equation}
\Psi(\beta_-,\Omega_-,\beta_+,\Omega_+)=e^{\mathrm{i}\sqrt{3}b_-\beta_-}K_{\mathrm{i}b_-}\left(\frac{\lambda}{\sqrt{3}}e^{-\sqrt{3}\Omega_-}\right)\,e^{\mathrm{i}\sqrt{3}b_+\beta_+}K_{\mathrm{i}b_+}\left(\frac{\mathrm{i}\lambda}{\sqrt{3}}e^{-\sqrt{3}\Omega_+}\right).
\end{equation}
The parameters $b_{\pm}$ can be regarded as momenta eigenvalues associated with the momentum operators $\hat{p}_{\beta_\pm}$ since $\hat{p}_{\beta_\pm}\Psi_{\pm}=\sqrt{3}b_\pm\,\Psi_{\pm}$. This implies that $b_\pm\in \mathbb{R}$ and admits continuous values. Therefore, the eigenfunctions for this problem are a family of Bessel functions, labeled by the eigenvalue of the associated $\beta$ momentum operator.
The squared modulus of the wave function is then written as
\begin{equation}
\left|\Psi \right|^2=K_{\mathrm{i}b_-}\left(\frac{\lambda}{\sqrt{3}}e^{-\sqrt{3}\Omega_-}\right)K^*_{\mathrm{i}b_-}\left(\frac{\lambda}{\sqrt{3}}e^{-\sqrt{3}\Omega_-}\right)\,K_{\mathrm{i}b_+}\left(\frac{\mathrm{i}\lambda}{\sqrt{3}}e^{-\sqrt{3}\Omega_+}\right)K^*_{\mathrm{i}b_+}\left(\frac{\mathrm{i}\lambda}{\sqrt{3}}e^{-\sqrt{3}\Omega_+}\right),
\end{equation}
which are depicted in Fig.~\ref{Psi_plus_and_minus}. The plot for $\left|\Psi\right|^2$ is given in Fig.~\ref{Psi_total}.

It is relevant to point out that $\left|\Psi\right|^2$ does not depend on $\beta$, which can be interpreted as a free wave parameter. However, according to the map, Eq.~(\ref{map}), $\beta$ is still relevant in distinguishing the point solutions from inner and outer spaces: i.e. any value for $\Omega$ is valid for both symmetrical points around the border between inner and outer spaces. Due to such a symmetrical behavior, for the physical limit given by $\Omega_\pm\rightarrow\infty$, one has the shell approaching either the singularity at $t=0$ or the event horizon at $t=2M$, which constrains the map to the interval $t \in [0,\,2M]$.
In particular, for $\Omega_\pm\rightarrow +\infty$, the wave function exhibits an oscillatory profile which can be interpreted as the absence of a meaningful limit for the quantization of the BH. Therefore no conclusions can be drawn from the limit corresponding to the BH singularity and event horizon. 
The vanishing of the wave function for $\Omega_\pm <0$ is consistent with this function having a minimum and therefore to exclude arbitrarily negative values for $\Omega_\pm$ for a fixed BH mass.

\subsection{Noncommutative scenario}

We consider now a deformed Heisenberg-Weyl algebra, where position and momentum do not commute \cite{Catarina_2008,Catarina_2010,Catarina_2010_2}:
\begin{equation} \label{NC_relations}
\left[\hat{\beta}_{NC},\hat{\Omega}_{NC}\right]=\mathrm{i}\theta, \qquad \left[\hat{\beta}_{NC},\hat{\pi}_\beta\right]=\mathrm{i}, \qquad \left[\hat{\Omega}_{NC},\hat{\pi}_\Omega\right]=\mathrm{i}, \qquad \left[\hat{\pi}_\beta,\hat{\pi}_\Omega\right]=\mathrm{i}\eta,
\end{equation}
where $\eta$ and $\theta$ are new constants of Nature. These variables can be mapped into commutative ones, i.e. those supported by standard quantum mechanics (Heisenberg-Weyl algebra) relations, through a noncanonical transformation usually referred to as Darboux or Seiberg-Witten map \cite{Seiberg_1999}. In particular, a useful transformation can be written as \cite{Catarina_2008}:
\begin{equation}
q_i=x_i-\frac{\theta_{ij}}{2}p_j, \qquad \pi_i=p_i+\frac{\eta_{ij}}{2}x_j,
\end{equation}
where $\theta_{ij}$ and $\eta_{ij}$ are antisymmetric matrices, which can be specialized to the BH problem in terms of the NC variables written as \cite{Catarina_2008,Catarina_2010,Catarina_2010_2}:
\begin{equation}
\hat{\beta}_{NC}=\hat{\beta}-\frac{\theta}{2}\hat{p}_{\Omega}, \quad \Omega_{NC}=\hat{\Omega}+\frac{\theta}{2}\hat{p}_{\beta},
\end{equation}
for the position, and \cite{Catarina_2008,Catarina_2010,Catarina_2010_2}
\begin{equation}
\hat{\pi}_\beta=\hat{p}_\beta+\frac{\eta}{2}\hat{\Omega}, \quad \hat{\pi}_\Omega=\hat{p}_\Omega-\frac{\eta}{2}\hat{\beta},
\end{equation}
for the momenta. With these transformations the Hamiltonian becomes:
\begin{equation} \label{NCHamiltonian}
\begin{aligned}
\hat{H}_{NC}=&\,e^{-2 \sqrt{3}\left(\hat{\Omega}_- +\frac{\theta}{2}\hat{p}_{\beta_-}\right)}\left[\left(\hat{p}_{\beta_-}+\frac{\eta }{2 } \hat{\Omega}_- \right)^2-\left(\hat{p}_{\Omega_-}-\frac{ \eta }{2 }\hat{\beta}_- \right)^2\right]+ \\
&+e^{-2 \sqrt{3}\left(\hat{\Omega}_+ +\frac{\theta}{2}\hat{p}_{\beta_+}\right)}\left[\left(\hat{p}_{\beta_+}+\frac{\eta }{2 } \hat{\Omega}_+ \right)^2-\left(\hat{p}_{\Omega_+}-\frac{ \eta }{2 }\hat{\beta}_+ \right)^2\right].
\end{aligned}
\end{equation}
At this point, a few remarks are in order. Notice first that this Hamiltonian is symmetric with respect to the exchange of the inner and outer metric variables. Furthermore, each term of the Hamiltonian depends only on one of the sets of metric variables, since there are no crossed terms of ``$+$'' and ``$-$'' variables. The WdW equation for the problem is thus the following:
\begin{equation}
\begin{aligned}
&\,e^{-2 \sqrt{3}\left(\hat{\Omega}_- -\mathrm{i}\frac{\theta}{2}\partial_{\beta_-}\right)}\left[\left(-\mathrm{i}\partial_{\beta_-}+\frac{\eta }{2 } \hat{\Omega}_- \right)^2-\left(-\mathrm{i}\partial_{\Omega_-}-\frac{ \eta }{2 }\hat{\beta}_- \right)^2\right]\Psi(\beta_-,\Omega_-,\beta_+,\Omega_+)+ \\
&+e^{-2 \sqrt{3}\left(\hat{\Omega}_+ -\mathrm{i}\frac{\theta}{2}\partial_{\beta_+}\right)}\left[\left(-\mathrm{i}\partial_{\beta_+}+\frac{\eta }{2 } \hat{\Omega}_+ \right)^2-\left(-\mathrm{i}\partial_{\Omega_+}-\frac{ \eta }{2 }\hat{\beta}_+ \right)^2\right]\Psi(\beta_-,\Omega_-,\beta_+,\Omega_+)=0,
\end{aligned}
\end{equation}
for which likewise the commutative case, the same decoupling features for the wavefunction are found.

Thus, we can write the decoupled eigenvalue equations:
\begin{subequations} \label{nc_eqs}
\begin{equation}
e^{-2 \sqrt{3}\left(\hat{\Omega}_- -\mathrm{i}\frac{\theta}{2}\partial_{\beta_-}\right)}\left[\left(-\mathrm{i}\partial_{\beta_-}+\frac{\eta }{2 } \hat{\Omega}_- \right)^2-\left(-\mathrm{i}\partial_{\Omega_-}-\frac{ \eta }{2 }\hat{\beta}_- \right)^2\right]\Psi_-(\beta_-,\Omega_-)=\lambda^2 \Psi_-(\beta_-,\Omega_-),
\end{equation}
\begin{equation}
e^{-2 \sqrt{3}\left(\hat{\Omega}_+ -\mathrm{i}\frac{\theta}{2}\partial_{\beta_+}\right)}\left[\left(-\mathrm{i}\partial_{\beta_+}+\frac{\eta }{2 } \hat{\Omega}_+ \right)^2-\left(-\mathrm{i}\partial_{\Omega_+}-\frac{ \eta }{2 }\hat{\beta}_+ \right)^2\right]\Psi_+(\beta_+,\Omega_+)=-\lambda^2\Psi_+(\beta_+,\Omega_+).
\end{equation}
\end{subequations}
In order to find the solution of the above equations, it is necessary to obtain a constant of motion associated with each of the separated Hamiltonians. It is straightforward to check that $\hat{B}_\pm=\hat{p}_{\beta_\pm}-\frac{\eta}{2}\hat{\Omega}_\pm$ commutes with $\hat{H}_\pm$ so that the solutions must obey the eigenvalue equations:
\begin{equation}
\hat{B}_\pm \Psi_\pm = \left(-\mathrm{i}\partial_{\beta_\pm}-\frac{\eta}{2}\hat{\Omega}_\pm\right)\Psi_\pm=b_\pm\Psi_\pm,
\end{equation} 
and thus the wave function can be written as
\begin{equation}
\Psi_\pm=e^{\frac{i \beta_\pm  }{2}(2 b_\pm +\eta  \Omega_\pm )}F_\pm(\Omega_\pm ).
\end{equation}
Substituting the above solutions into Eqs.~(\ref{nc_eqs}) leads to ordinary differential equations for the functions $F_\pm$:
\begin{equation} \label{full_NC_Omega}
F(\Omega ) \left(b_\pm^2 +\eta ^2 \Omega ^2 +2 b_\pm \eta  \Omega  \mp e^{-\frac{\sqrt{3}}{2} (2\theta b_{\pm}  +\eta  \theta  \Omega +4 \Omega )}\lambda^2 \right)+F''(\Omega ) =0.
\end{equation}

\begin{figure}[h]
    \includegraphics[width=0.8\textwidth]{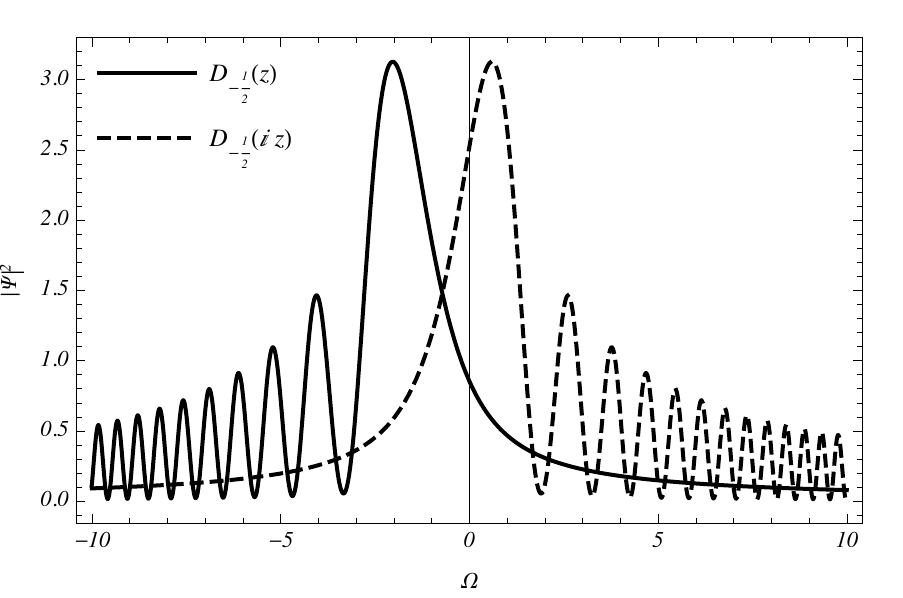}
    \caption{$\left|D_{-1/2}(z)\right|^2$ and $\left|D_{-1/2}(\mathrm{i}z)\right|^2$ as functions of $\Omega_\pm$. The parameters are $b_{\pm} = 0.5$ and $\eta=0.7$. Here $z=\frac{(1+i) (b_{\pm}+\eta  \Omega )}{\sqrt{\eta }}$.}
    \label{Psi_plus_and_minus_NC}
\end{figure}

Exact solutions for such equations are difficult to find for $\lambda\neq 0$. Therefore, we search for numerical solutions that satisfy the required boundary conditions. For the particular choice of $\lambda=0$ an exact solution can be found for the ordinary differential equation:
\begin{equation}
F(\Omega_\pm ) (b_\pm  +\eta\,\Omega_\pm )^2+ F''(\Omega_\pm )=0,
\end{equation}
that is
\begin{equation}
F_\pm(\Omega_\pm)=c_1\,D_{-\frac{1}{2}}\left(\frac{(1+i) (b_\pm  +\eta\,\Omega_\pm )}{\sqrt{\eta }  }\right)+c_2\,D_{-\frac{1}{2}}\left(-\frac{(1-i) (b_\pm  +\eta\,\Omega_\pm )}{\sqrt{\eta }  }\right),
\end{equation}
where $D_\nu(z)$ are parabolic cylinder functions, depicted in Fig.~\ref{Psi_plus_and_minus_NC}.
They correspond to mirror reflected solutions with a damped oscillatory profile that vanish at $\Omega\rightarrow\pm\infty$.
This feature arises from the introduction of the noncommutativity, more specifically, due to the momentum noncommutativity, since the parameter $\theta$ is absent from the WdW equation for $\lambda=0$.

As previously argued, we should focus only on the behavior of the functions for $\Omega > 0$. One can see that, at least for $\lambda=0$, the probability density in the limit $t\rightarrow 0$ vanishes. Furthermore, since the dependence of $\Psi$ on $\beta_\pm$ appears only as a phase factor, this dependance disappears for $\left|\Psi\right|^2$. Thus, it is possible to conclude that the probability density in the limit $t\rightarrow 2M$ also vanishes. Since both geometries on either side of the shell have a vanishing probability density to reach the singularity, then, the shell, which is defined as the separation surface between these geometries, also has a vanishing probability density of collapsing into the singularity. The same arguments applies for $t=2M$: the shell can never leave the event horizon. 

\begin{figure}[h]
    \includegraphics[width=0.6\textwidth]{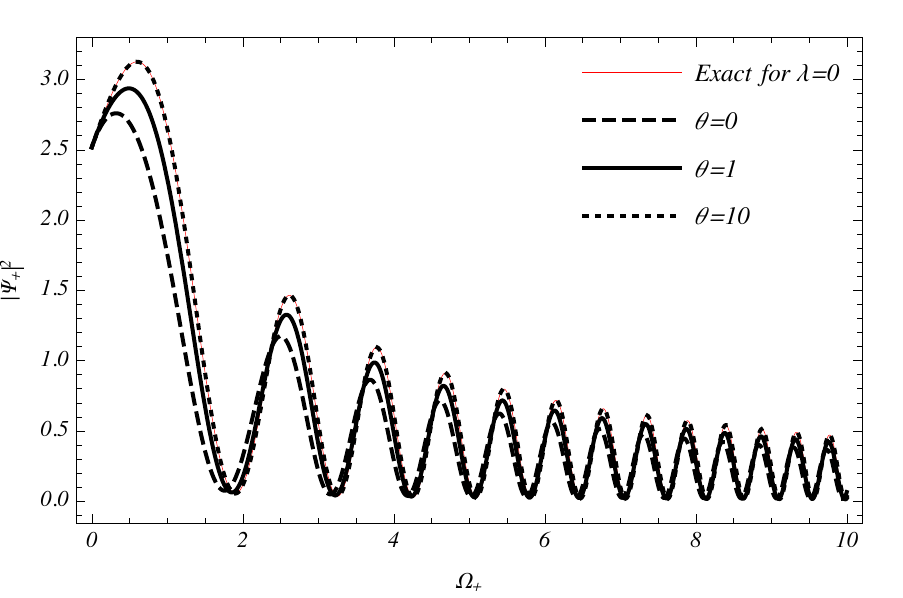}
    \includegraphics[width=0.6\textwidth]{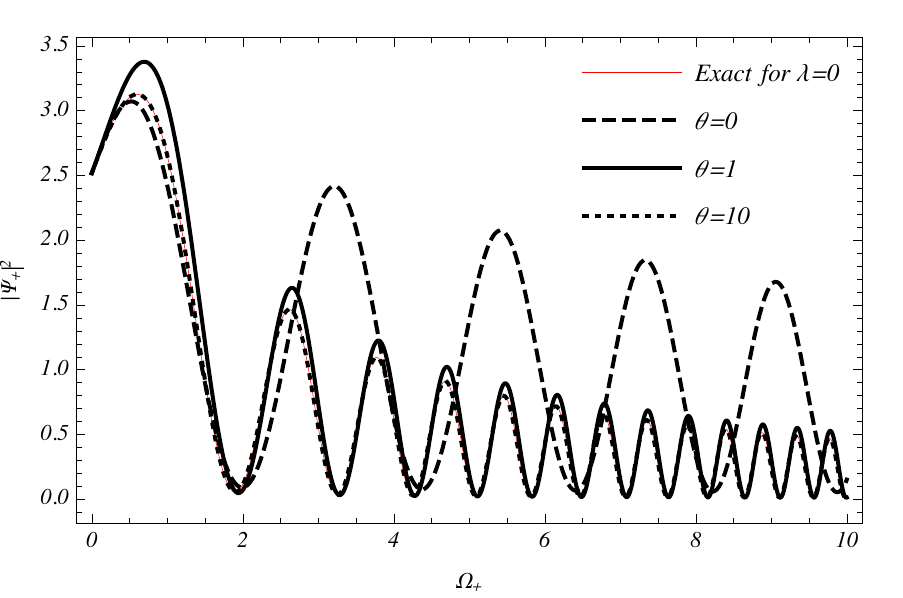}
    \caption{Numerical solutions of Eq.~(\ref{full_NC_Omega}) for $\lambda = 1$ as function of $\Omega_-$ (first plot) and $\Omega_+$ (second plot), with $b_\pm =0.5$ and $\eta=0.7$. 
    Solutions are for $\theta = 0$ (dashed lines), $\theta = 1$ (solid lines), and $\theta = 10$ (dotted lines).}
    \label{Psi_neg_lambda}
\end{figure}

Although the exact solution discussed above portrays the general characteristics of the solutions when noncommutativity is introduced, one must study the numerical solutions of Eq.~(\ref{full_NC_Omega}) for $\lambda\neq 0$ since, for this cases, the effect of $\theta$ no longer disappears. By consistency, the boundary conditions to be used are the ones from the exact solutions for $\lambda=0$, with $b_\pm = 0.5$ and $\eta=0.7$. 
As expected, the choice of boundary conditions has no significative effect on the behavior of the wave functions, in particular, in the limit of interest, $\Omega\rightarrow\infty$. This can be better understood through the analysis of the asymptotic behavior of Eq.~(\ref{full_NC_Omega}), which gives rise to:
\begin{equation}
F(\Omega_\pm ) (b_\pm  +\eta\,\Omega_\pm )^2+ F''(\Omega_\pm )=0,
\end{equation}
i.e. the same equation for the $\lambda=0$ scenario. The effect of $\lambda$ is relevant only nearby $\Omega=0$. The numerical results for some non-vanishing values of $\lambda$ and $\theta$ are shown in Fig. \ref{Psi_neg_lambda}.


It can be seen that the damping behavior of the function found for the $\lambda=0$ case is maintained. Furthermore, we find that damping is clearly a feature due to the momentum noncommutativity as it still remains for $\theta=0$. Consequently, the conclusions drawn from the exact solution can be extended to the numerical ones. Namely, the vanishing of the probability density at the singularity and the event horizon at $t=2M$. In order to compute the probability of the wave function, one must integrate over $\Omega_\pm$ on a surface of constant $\beta_\pm$. The integration measure is $\delta(\beta-\beta_c)\mathrm{d}\beta\,\mathrm{d}\Omega$ \cite{Catarina_2008}, which leads to:
\begin{equation}
P(\beta,\Omega)=\int \delta(\beta'-\beta_c)\mathrm{d}\beta'\,\mathrm{d}\Omega' \left|F(\Omega')\right|^2\sim \int_0^{+\infty} \mathrm{d}\Omega' \left|F(\Omega')\right|^2.
\end{equation}
Despite the vanishing of the wave function for $\Omega\rightarrow\infty$, the above integral is not convergent, so the obtained solution is not squared integrable. This is a feature also encountered for the NC black hole without a shell \cite{Catarina_2010}, for the commutation relations Eqs.~(\ref{NC_relations}).

In any case, the effect of $\eta$ alone leads to the damping of the wave function that is responsible for its regularization at the limit $\Omega\rightarrow\infty$. It is easy to see that the damping disapears for $\eta=0$. Indeed considering Eq.~(\ref{full_NC_Omega}) and in the limit $\eta\rightarrow 0$:
\begin{equation}
F(\Omega ) \left(b_\pm^2 \mp e^{-\frac{1}{2} \sqrt{3} (2 \theta b_\pm +4 \Omega )}\lambda \right)+F''(\Omega) =0.
\end{equation}
This equation admits exact solutions similar to the ones in Eq.~(\ref{Commutative_sols}), so that the complete wave functions $\Psi_\pm$ can be written as:
\begin{subequations} \label{theta_effect}
\begin{equation}
\Psi_{-}(\beta_-,\Omega_-)=e^{\mathrm{i}\sqrt{3}b_-\beta_-}K_{\mathrm{i}b_-}\left(\frac{\lambda}{\sqrt{3}}e^{-\frac{\sqrt{3}}{2}\left(\theta b_- +2\Omega_-\right)}\right),
\end{equation}
\begin{equation}
\Psi_{+}(\beta_+,\Omega_+)=e^{\mathrm{i}\sqrt{3}b_+\beta_+}K_{\mathrm{i}b_+}\left(\frac{\mathrm{i}\lambda}{\sqrt{3}}e^{-\frac{\sqrt{3}}{2}\left(\theta b_+ +2\Omega_+\right)}\right),
\end{equation}
\end{subequations}
which corresponds to a shift in the oscilatory behaviour of the wave function, Eqs.~(\ref{Commutative_sols}).

\section{Conclusions} \label{conclusions_section}

In this work, a spherically symmetric null-like hypersurface collapsing into a BH is considered. The classical and quantum framework for the problem are constructed through the assumptions that inside and outside regions of the shell are mapped into two distinct KS metrics. The junction condition formalism for null hypersurfaces is applied in order to obtain the action for the system. Furthermore, since this hypersurface is a null boundary for both KS spacetimes, an extra boundary term for the action is admited \cite{Parattu_2016,Lehner_2016}. This allows for the definition of the Hamiltonian of the system and for obtaining the WdW equation for the quantum mechanical problem.

The WdW equation is solved, and the wavefunction is found to be given by a family of modified Bessel functions, which exhibit an oscillatory behavior in the limit corresponding to the BH singularity, $\Omega\rightarrow\infty$. The squared modulus of this wave function does not depend on $\beta$, thus, from the map, Eq.~(\ref{map}), there is a symmetry with respect to the raduis $t=M$. Therefore, the limit $\Omega\rightarrow\infty$ corresponds to both the singularity and the event horizon.

Extending the Heisenberg-Weyl algebra and considering the phase space noncommutativity, it is shown that the configuration space $\theta$ parameter contributes only as a shift in the wave function of the commutative problem. On the other hand, when the NC parameter $\eta$ is considered, a damped wave function arises for $\Omega\rightarrow\infty$. It is found that the squared modulus of the wave function does not depend on $\beta$, thus the probability density vanishes as $t\rightarrow 0$ and $t\rightarrow 2M$. Thus, the introduction of the momentum noncommutativity acts as a regulator for the wave function, a feature previously encountered for the BH problem without the shell \cite{Catarina_2010}. This result suggests that a more complex NC framework is required to obtain square integrable wave functions that vanish at the singularity as found in Ref.~\cite{Catarina_2010_2}.

\section*{Appendix -- Vanishing of the surface term for different time coordinates \label{appendices}}

In Sec.~\ref{action_section} it was argued that one could take $t$ the coordinate to be the same on both parts of the manifold. However, this is not necessarily true and the same result, namely the vanishing of the boundary terms, can be recovered by assuming that the time coordinates are different.

Indeed, starting from metrics of the form:
\begin{equation}
\mathrm{d}s^2_\pm=e^{\sqrt{3}\beta\pm}\,\mathrm{d}u\left(e^{\sqrt{3}\beta_\pm}\,\mathrm{d}u+2\zeta N\pm\,\mathrm{d}t_\pm\right)+e^{-2\sqrt{3}\Lambda}\,\left(\mathrm{d}\theta^2+\sin^2\theta\,\mathrm{d}\phi^2\right),
\end{equation}
one can perform a change of coordinates, $T_\pm\defeq e^{-2\sqrt{3}\Lambda_\pm}$, so to obtain:
\begin{equation}
\mathrm{d}s^2_\pm=e^{\sqrt{3}\beta_\pm}\,\mathrm{d}u\left(e^{\sqrt{3}\beta_\pm}\,\mathrm{d}u-\frac{2\zeta N}{\sqrt{3}T_\pm\,\alpha_\pm}\,\mathrm{d}T_\pm\right)+T_\pm^2\,\left(\mathrm{d}\theta^2+\sin^2\theta\,\mathrm{d}\phi^2\right),
\end{equation}
where $\alpha_\pm(T)=\dot{\Lambda}_\pm(T)$. The holonomic vectors are then given by:
\begin{equation}
e^\mu_{(A)}\defeq \frac{\partial x^\mu}{\partial y^A}=\delta^\mu_A,
\end{equation}
for $A=\theta,\phi$, and by
\begin{equation} \label{new_coord}
e^\mu_{(T)}\defeq \frac{\partial x^\mu}{\partial T}=\partial_T,
\end{equation}
for $T_\pm$.
These definitions allow for the computation of the induced metric using $g_{ab}=g_{\mu\nu}e^\mu_{(a)}e^\nu_{(b)}$ for both sides as:
\begin{equation} 
\mathrm{d}s^2_{|_\Sigma}=T_\pm^2\,\mathrm{d}\Omega^2. 
\end{equation}
For the hypersurface to possess a well defined structure, the induced metrics on each of its sides must coincide, and the coordinates $T_\pm$ must be the same. In the coordinate system considered in the body of the paper (with $t$ rather than $T$), this leads to the condition:
\begin{equation}
\Lambda_+(t_+)=\Lambda_-(t_-).
\end{equation}
Besides providing the condition for the junction of the spacetimes, this coordinate system also allows for a convenient set of coordinates to be introduced in the shell, the coordinates $\left(T,\theta,\phi\right)$. These are the natural coordinates on $\Sigma$ and they are the same on both sides of the hypersurface.
Using the calculations performed in Sec.~\ref{action_section} alone, in the novel coordinate system, Eq. (\ref{new_coord}), the normal and transversal vectors to $\Sigma$ are then given by:
\begin{equation}
n_\pm^\mu=-\frac{\zeta \, T\, \alpha _\pm e^{-\sqrt{3} \beta _\pm}}{N_\pm \chi _\pm}\,\partial_T,
\end{equation}
\begin{equation}
M^\mu_\pm=\frac{\chi _\pm}{\varepsilon_\pm}\partial_u-\frac{\zeta \, T \,\alpha _\pm e^{\sqrt{3} \beta_\pm} \chi _\pm}{2 \varepsilon_\pm N_\pm}\,\partial_T.
\end{equation}
The projections of these vectores onto the three dimensional hypersurface are:
\begin{equation}
n^a_\pm=-\frac{\zeta \, T\, \alpha _\pm e^{-\sqrt{3} \beta _\pm}}{N_\pm \chi _\pm}\,\partial_T,
\end{equation}
for the normal vector, and
\begin{equation}
M_a^\pm=-\frac{\zeta  N_\pm e^{\sqrt{3} \beta _\pm} \chi _\pm}{\varepsilon_\pm \,T \,\alpha _\pm}\,\partial_T,
\end{equation}
for the transverse one.
The requirement that the quantities $n^a$ and $M_a$ coincide on both sides of the shell determines a unique structure for $\Sigma$. This condition imposes restrictions on the normalization factors, namely:
\begin{equation} \label{boundaryCondition2}
\frac{   e^{-\sqrt{3} \beta _-} \alpha_-}{N_- \chi _-}\Big|_{u=u_0}=\frac{   e^{-\sqrt{3} \beta _+} \alpha_+}{N_+ \chi_+}\Big|_{u=u_0},
\end{equation}
and,
\begin{equation}
\frac{  N_- e^{\sqrt{3} \beta _-} \chi _-}{ \varepsilon _- \alpha_-}\Big|_{u=u_0}=\frac{  N_+ e^{\sqrt{3} \beta _+}\chi _+}{ \varepsilon _+ \alpha_+}\Big|_{u=u_0}.
\end{equation}
Once again, these lead to the conclusion that $\varepsilon_-=\varepsilon_+$. The next step is to compute the transverse curvature, which is found to be,
\begin{equation}
\begin{aligned}
\mathcal{K}_{tt}^\pm&=\frac{\zeta  e^{\sqrt{3} \beta _\pm} \chi _\pm}{\varepsilon \, T^2\, \alpha _\pm^2} \left[T \alpha _\pm \dot{N}_\pm+N_\pm \left(\alpha _\pm \left(\sqrt{3} T \dot{\beta} _\pm-1\right)-T \dot{\alpha} _\pm\right)\right], \\
\mathcal{K}_{\theta\theta}^\pm&=\frac{\zeta \, T^2\, \alpha _\pm e^{\sqrt{3} \beta _\pm} \chi _\pm}{2 \varepsilon N_\pm}, \\
\mathcal{K}_{\phi\phi}^\pm&=\mathcal{K}_{\theta\theta}^\pm\sin ^2(\theta ),
\end{aligned}
\end{equation}
as well as the discontinuity in the curvature scalar,
\begin{equation}
\begin{aligned}
R_{|_\Sigma}=-\frac{2 \zeta  e^{-\sqrt{3} \beta _-}}{\varepsilon  N_-^2 \chi _-} \left[N_- \left(\alpha _- \left(\sqrt{3} T \dot{\beta} _--1\right)-T \dot{\alpha} _-\right)+\alpha _- T \dot{N}_-\right]+ \\
+\frac{2 \zeta  e^{-\sqrt{3} \beta _+}}{\varepsilon  N_+^2 \chi _+} \left[N_+ \left(\alpha _+ \left(\sqrt{3} T \dot{\beta} _+-1\right)-T \dot{\alpha} _+\right)+\alpha _+ T \dot{N}_+\right],
\end{aligned}
\end{equation}
where Eq.~(\ref{boundaryCondition2}) was used in order to organize the expression.
Aiming to compute the boundary action, we are left with the task of computing the trace of the $\Theta$ tensor and the non affinity parameter $\kappa$. In this setup their respective expressions are the following:
\begin{equation}
\Theta_\pm=-\frac{2 \zeta  \alpha _\pm e^{-\sqrt{3} \beta _\pm}}{N_\pm \chi _\pm}, \qquad \kappa_\pm=\frac{\zeta \, T \,\alpha _\pm e^{-\sqrt{3} \beta _\pm} \dot{\chi} _\pm}{N_\pm \chi _\pm^2}.
\end{equation}
Finally, once these results are inserted into the piece of the Lagrangian corresponding to the boundary, $\Sigma$, it vanishes and the boundary terms in the action are suppressed. Thus, the result of Sec.~\ref{action_section} remains unchanged even under more general assumptions.

\vspace{.5 cm}
{\em Acknowledgments} -- The work of PL is supported by FCT (Portuguese Funda\c{c}\~ao para Ci\^encia e Tecnologia) grant PD/BD/135005/2017. The work of AEB is supported by the Brazilian agency FAPESP (grant 17/02294-2). The work of OB is partially supported by the COST action MP1405.

\end{document}